\documentclass[letterpaper, 10pt, conference]{IEEEtran} 

\IEEEoverridecommandlockouts
\usepackage{amsmath} 
\usepackage{graphicx} 
\usepackage{booktabs} 
\usepackage{siunitx} 
\usepackage{amssymb}
\usepackage{color}

\begin{document}



\title{Spatial Association Between Near-Misses and Accident Blackspots in Sydney, Australia: A Getis-Ord $G_i^*$ Analysis}

\author{
\IEEEauthorblockN{1\textsuperscript{st} \textbf{Artur Grigorev}}
\IEEEauthorblockA{\textit{Faculty of Engineering and IT} \\
\textit{University of Technology Sydney}\\
Sydney, Australia \\
ORCID: 0000-0001-6875-3568}
\and 
\IEEEauthorblockN{2\textsuperscript{nd} \textbf{David Lillo-Trynes}}
\IEEEauthorblockA{
\textit{Compass IoT}\\                   
Sydney, Australia                        
}
\and 
\IEEEauthorblockN{3\textsuperscript{rd} \textbf{Adriana-Simona~Mih\u{a}i\c{t}\u{a}}} 
\IEEEauthorblockA{\textit{Faculty of Engineering and IT} \\
\textit{University of Technology Sydney}\\
Sydney, Australia \\
ORCID: 0000-0001-7670-5777}
}
\maketitle

\begin{abstract} 

Conventional road safety management is inherently reactive, relying on analysis of sparse and lagged historical crash data to identify hazardous locations, or crash blackspots. The proliferation of vehicle telematics presents an opportunity for a paradigm shift towards proactive safety, using high-frequency, high-resolution near-miss data as a leading indicator of crash risk. This paper presents a spatial-statistical framework to systematically analyze the concordance and discordance between official crash records and near-miss events within urban environment. 
A Getis-Ord statistic is first applied to both reported crashes and near-miss events to identify statistically significant local clusters of each type. Subsequently, Bivariate Local Moran's I assesses spatial relationships between crash counts and High-G event counts, classifying grid cells into distinct profiles: High-High (coincident risk), High-Low and Low-High. Our analysis reveals significant amount of Low-Crash, High-Near-Miss clusters representing high-risk areas that remain unobservable when relying solely on historical crash data. Feature importance analysis is performed using contextual Point of Interest data to identify the different infrastructure factors that characterize difference between spatial clusters. The results provide a data-driven methodology for transport authorities to transition from a reactive to a proactive safety management strategy, allowing targeted interventions before severe crashes occur.

\end{abstract}


\section{Introduction}

Road traffic accidents impose significant societal costs globally, demanding effective safety management strategies. Historically, identifying high-risk locations has predominantly relied on analysing police-reported accident data (A). While valuable, this approach is reactive, requiring accidents to occur before interventions are typically considered. Furthermore, accident data, particularly for severe incidents, can be sparse in space and time, making statistical identification of hazardous locations challenging \cite{PMC7070501}.

The emergence of vehicle telematics and advanced driver-assistance systems (ADAS) has enabled the collection of vast amounts of data on driver behaviour and vehicle kinematics, including near-miss events \cite{MnDOT_WorkZoneNDS}. Near-misses, defined as unsafe events where a collision is narrowly avoided, occur much more frequently than actual accidents and are considered leading indicators of underlying safety risks \cite{TarkoSSMReview2018, MDPI_SSM_Review}. Analysing near-miss patterns offers a proactive approach to identifying potentially hazardous locations before serious accidents occur.

Within the spectrum of near-miss data, events characterised by high severity metrics – such as large longitudinal or lateral accelerations – are hypothesised to represent situations with a higher potential for resulting in injury should a collision occur \cite{ResearchGateSSMCompareEVT}. We further define this as G-Force, a measure of the acceleration during a nearmiss event (and later on as NM+G a nearmiss with a high G-force).
We believe that by utilizing NM+G severity data, we can proactively refine the safety analysis by focusing on locations where the ``potential consequences'' of frequent conflicts are much higher.


\textbf{Gap:} Despite the increasing availability of near-miss frequency data, the spatial relationship between hotspots derived from \textit{near-miss severity} indicators (such as high G-force events, hereafter High-G) and hotspots identified using traditional \textit{accident} data (A) remains relatively unexplored, particularly at fine spatial resolutions within urban areas. To the best of our knowledge, it has never been proved whether areas exhibiting frequent high-severity near-miss events (NM+G) directly correspond spatially to areas where accidents are historically concentrated, or if NM+G events will significantly increase the occurrence of actual crashes. Establishing this link is critical for validating the use of High-G severity as a reliable spatial proxy for accident risk, enabling a more effective proactive safety management, and revealing different dimensions of underlying risk. Significant spatial discordance, for instance, might indicate areas with a high latent risk undetected by sparse accident data, or conversely, areas where underlying factors successfully mitigate the consequences of frequent severe near-misses.

\textbf{Objective:} Therefore, this research aims to evaluate the impact of high-severity near-miss events (High-G) on blackspot areas (A) already mapped by past historical traffic incident logs. For This study we use data provided by Compass IoT, a leading Australian startup collecting real-time data from Connected Vehicles across Australia, with a dedicated focus on Sydney, the largest city with the highest traffic incident levels. We use a 400m grid framework established for the study period (2022), and we quantitatively characterise and correlate these lagging and leading safety indicators. The specific objectives are to:
\begin{enumerate}
    \item Identify statistically significant spatial clusters (hotspots/coldspots) of reported accidents (A) via the Getis-Ord $G_i^*$ statistic on the 400m grid.
    \item Aggregate High-G event data onto the same 400m grid framework for fine resolution mapping. 
    \item Quantify and map the local spatial correlation between aggregated crash and High-G counts using Bivariate Moran's I (LISA) to identify distinct concordance/discordance patterns (HH, LL, HL, LH).
    \item Characterise the identified LISA pattern areas using Point of Interest (POI) counts and assess the contribution of these environmental features in differentiating key spatial risk profiles (e.g., HH vs. LL). 
\end{enumerate}
This evaluation of local spatial correlation patterns (via LISA) between leading (High-G) and lagging (A) indicators provides a nuanced understanding of potential versus realized traffic risk across diverse urban settings. Characterizing these patterns with POI data further elucidates the limited role of such static environmental context in explaining risk variations. The findings directly inform the practical application and limitations of using telematics-derived severity data for network screening and targeted safety interventions.

\section{Related Works}

Road safety analysis has traditionally centered on reactive approaches, primarily identifying high-risk locations, often termed hotspots or blackspots, based on historical police-reported accident data \cite{PMC7070501, PMC10878868}. Methodologies evolved from simple frequency rankings or accident rate calculations to more sophisticated spatial statistical techniques deployed within Geographic Information Systems (GIS) \cite{PMC5251886}. Prominent methods include Kernel Density Estimation (KDE) for visualizing density \cite{MDPI_SpatialTemporalAnalysis} and spatial autocorrelation analyses, such as Moran's I for assessing global clustering \cite{ColumbiaHotspotReview} and Local Indicators of Spatial Association (LISA) like Getis-Ord $G_i^*$ and Local Moran's I for pinpointing statistically significant local clusters of high (hotspots) or low (coldspots) incident counts \cite{ColumbiaHotspotReview, Anselin1995, ProQuestGISHotspots}. However, the fundamental limitations of this approach are well-documented: its reactive nature (requiring crashes to occur first), the relative rarity and potential underreporting or inaccuracy of official crash data, and the ethical concerns associated with waiting for harm \cite{PMC7070501, TandF_SSM_Heterogeneous, TandF_SSM_Curves}.

These limitations inspire a significant shift towards proactive methodologies using Surrogate Safety Measures (SSMs) \cite{TandF_SSM_Heterogeneous}. The foundation lies in the Traffic Conflict Technique (TCT) to systematically observe near-miss events \cite{TarkoSSMReview2018, ShafizadehTCTReview}. A near-miss or traffic conflict is generally defined as an interaction necessitating an evasive maneuver to avoid a collision \cite{TandF_SSM_Heterogeneous, TarkoSSMReview2018}. These non-crash events occur far more frequently than actual crashes, providing statistically richer datasets for analyzing underlying traffic risks and evaluating countermeasures without relying on sparse crash data \cite{TandF_SSM_Heterogeneous, MDPI_SSM_Review}. Consequently, research has focused on developing and applying various SSM indicators derived from detailed observational or sensor data. Common indicators include temporal measures like Time-to-Collision (TTC) and Post-Encroachment Time (PET), deceleration requirements like DRAC, and kinematic indicators such as speed, lateral deviation, or harsh events (e.g., rapid braking/acceleration) \cite{TandF_SSM_Heterogeneous, MDPI_SSM_Review}. Data for SSM calculation is usually sourced from video analytics platforms using computer vision and AI \cite{PMC10943440_ConflictEval, MDPI_ITS_Edge, ResearchGateVideoAnalytics}, instrumented vehicles in Naturalistic Driving Studies (NDS) \cite{MnDOT_WorkZoneNDS, NAS_SHRP2_RID}, smartphone sensors \cite{MDPI_SSM_Review}, and Connected Vehicle (CV) data streams \cite{MDPI_SSM_Review, ResearchGateCVHive}.

Analyzing crash causation to develop effective safety countermeasures is essential, yet hindered by the low frequency of actual crash events, particularly within rich datasets from Naturalistic Driving Studies (NDS) \cite{Wang2022Causation}. NDS provides detailed real-world driving data but typically captures fewer crashes relative to the volume of driving. This data scarcity necessitates the use of SSMs - observable events, like traffic conflicts or near-crashes, thought to be correlated with crash risk \cite{MDPI_SSM_Review}. However, establishing the validity of SSMs as reliable proxies for crash risk remains a significant and ongoing challenge in traffic safety research \cite{CSP_TCT_OpenQuestions}.

Near-crashes, commonly defined as events requiring a rapid evasive maneuver to avoid a collision, are frequently used SSMs, especially in NDS analysis. Their application often relies on the ``causal continuum'' hypothesis, which states that near-crashes and crashes arise from largely similar or identical causal factors. This assumption is considered plausible and supported by reported correlations between conflict/near-crash frequency and historical crash rates \cite{MDPI_SSM_Review}. Foundational research by Guo et al. \cite{guo2010near} provided critical support for this approach by demonstrating that near-crashes often exhibit kinematic signatures similar to crashes and appear to share common underlying causal mechanisms or contributing factors. These findings established near-crashes as viable and effective surrogates, allowing researchers to analyze the more abundant near-crash data to understand risk and causation.

Extreme Value Theory (EVT) offers a promising statistical framework to formally link the distribution of frequent surrogate events to the probability of rare, extreme crash events \cite{TandF_SSM_Heterogeneous, ResearchGateSSMCompareEVT, ResearchGateFreewayEVT}, though its application requires careful consideration of assumptions and data quality \cite{ResearchGateSSMCompareEVT}. Furthermore, the effectiveness and interpretation of SSMs are highly context-dependent, influenced by factors like traffic composition (homogeneous vs. heterogeneous), road geometry (intersections, curves), and environmental conditions, necessitating context-specific indicator selection and threshold calibration \cite{TandF_SSM_Heterogeneous, TandF_SSM_Curves, TandF_TrajectoryTransection}. Spatial analysis of near-miss frequency hotspots has become an area of growing interest for identifying general conflict-prone areas \cite{ResearchGateVideoAnalytics}.

While near-miss frequency indicates the prevalence of conflicts, metrics reflecting near-miss severity, such as high G-force events (denoted as NM+G) captured by inertial sensors in vehicles or smartphones, offer outlook into the potential consequence or danger level of these interactions \cite{MDPI_SSM_Review}. The $G-force$ values for emergency braking
vary significantly across multiple studies \cite{SAMSON2022126} with deceleration rates of vehicles found to be ranged from 0.49 m/s2 to 8.76 m/s2 with a total weighted average of 2.82 m/s2. In practice, some Automated Emergency Braking systems consider the threshold of 0.47g requiring automated intervention to prevent a collision: Bendix system uses 0.47g as the critical threshold to determine if a driver's braking is insufficient \cite{DOTHS812390}. The fleet telematics solution Geotab uses 0.47g as a pre-defined setting for flagging harsh cornering events \cite{Broughall_2020}. The threshold of 0.47g is selected to select near miss trajectories for this study.

High G-force readings, which signify rapid deceleration or acceleration, capture the severe vehicle maneuvers (near-miss events). Consequently, they have the potential to act as a robust proxy for identifying locations with a high risk of future collisions. However, research directly investigating the relationship between the severity of surrogate events and the severity of actual crashes is less developed than frequency-based comparisons. Specifically, there is limited work that directly compares the spatial patterns of hotspots derived from \textit{near-miss severity indicators} (like NM+G) with hotspots identified using crash data. This study aims to address this specific gap by using established spatial statistical methods (Getis-Ord $G_i^*$) to explicitly examine the concordance and discordance between hotspots identified using high G-force near-miss data and those identified using historical crash data, stratified by crash severity. This comparison seeks to clarify the utility of near-miss data as a spatial proxy for accident risk.

\section{Case study}
\label{sec:case}
This study integrates three key spatial datasets for the Sydney Greater Metropolitan Area, focusing on the year 2022:

\begin{itemize}
    \item \textbf{Road Accidents:} Locations of reported traffic crashes extracted for the year 2022 from a dataset (originally covering Jan 2017 - Jul 2022) sourced from Transport for New South Wales (TfNSW) (n = 3,658 points).
    \item \textbf{High G-Force Near Misses:} Locations identified from vehicle trajectory data provided by Compass IoT for 2022. These represent the point of maximum G-force recorded within vehicle trajectories where a near-miss event was detected (n = 24,137 points).
    \item \textbf{Points of Interest (POIs):} Geographic locations of various amenities, shops, and other features extracted from OpenStreetMap (OSM) data.
\end{itemize}

\section{Methodology}
\label{sec:methodology}

The instantaneous G-force quantifies the acceleration experienced relative to standard gravitational acceleration $ g = 9.806 \, \mathrm{m/s}^2 $. Let $\mathbf{a}(t) = [a_x(t), a_y(t), a_z(t)]^\top \in \mathbb{R}^3$ denote the time-dependent vector of longitudinal ($ a_x $), lateral ($ a_y $), and vertical ($ a_z $) accelerations measured by inertial sensors. The Euclidean norm $\left\| \cdot \right\|$ is computed as:  
$$
\left\| \mathbf{a}(t) \right\| = \sqrt{a_x(t)^2 + a_y(t)^2 + a_z(t)^2}.
$$  
The G-force at time $ t $ is defined via the ratio:  
$$
\text{G-force}(t) := \frac{\left\| \mathbf{a}(t) \right\|}{g} = \frac{\sqrt{a_x(t)^2 + a_y(t)^2 + a_z(t)^2}}{9.806}.
$$

The trajectory of both the the steering (lateral deviations) and braking (longitudinal deceleration) maneuvers exhibit localized extrema of G force (Fig. \ref{fig:g_force_trajectories}). 
Since both types of events have peak kinematic extrema localized within narrow time window, near miss severity can be spatially represented through discrete point of maximum G force within near miss trajectory.

While metrics like Time-to-Collision (TTC) or Post-Encroachment Time (PET) focus on temporal proximity during observed conflicts, this study prioritizes G-force as a kinematic proxy due to critical data limitations. With connected vehicle coverage limited to a small fraction of Sydney’s traffic flow and no access to multi-vehicle collision trajectories, our data is restricted to near-miss events from single vehicles. Furthermore, TTC and PET, as metrics of interaction, require the determination of a conflict point, which is not observable from single-vehicle data.

The core of this study involved a quantitative spatial analysis of reported crashes (A, $n=\num{3638}$) and high G-force events (High-G, $n=\num{24137}$) within Sydney (mapped for year 2022), using data aggregated onto a uniform 400m grid. The methodology focused on identifying statistically significant crash hotspots, analyzing the local spatial correlation between crashes and High-G events, and characterizing the resulting spatial patterns using POI data. Key stages that have been applied are the following:

\begin{figure}[htbp]
    \centering
    \includegraphics[width=0.9\linewidth]{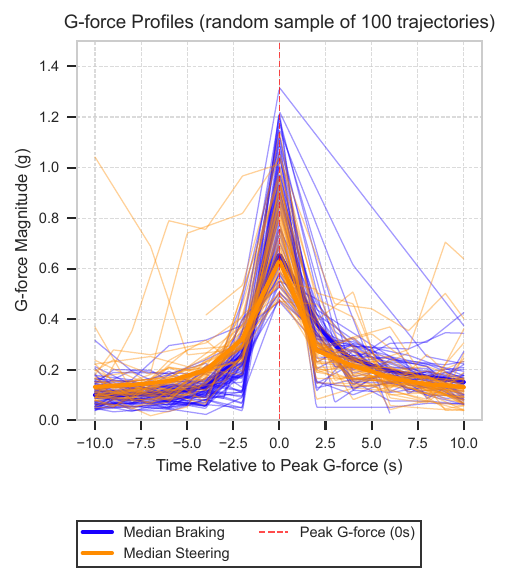}
    
    \caption{
        Vehicle trajectory segments represented with G-force profiles (x-axis: relative time centered around point of max G force; y-axis: acceleration in $g$)
    }
    \label{fig:g_force_trajectories}
\end{figure}

To facilitate our area-based analysis, a uniform 400m\,x\,400m grid was established across the study region (Projected CRS: EPSG:32756), resulting in \num{38824} cells. The size of the grid has been selected after conducting a sensitivity analysis of the best grid size to provide granular view into risky maneuvers. Data of the reported crashes and High-G events were spatially aggregated onto this grid, yielding cell-level counts for each variable (\textit{crash\_count}, \textit{highg\_count}). These aggregated counts formed the primary input for the spatial statistical analyses described below. 


Spatial clustering and correlation were assessed using established geospatial statistics applied to the aggregated grid data:
\begin{itemize}
    \item \textbf{Defining Spatial Relationships:} A Spatial Weights Matrix (SWM) based on Queen contiguity was constructed to formally define the neighborhood structure and spatial influence between adjacent grid cells.
    \item \textbf{Crash Hotspot Detection (Gi*):} The Getis-Ord $G_i^*$ statistic, a local indicator of spatial association, was calculated for each grid cell based solely on its \textit{crash\_count} relative to its neighbors (defined by the SWM). Significance testing ($p<0.10$, $p<0.05$, $p<0.01$) using permutation inference identified statistically significant crash \textit{hot spots} (high-crash clusters) and \textit{cold spots} (low-crash clusters). 
    \item \textbf{Bivariate Spatial Correlation (LISA):} LISA analysis was used to quantify and map the local spatial correlation between \textit{crash\_count} and \textit{highg\_count}. This identified cells exhibiting statistically significant ($p<0.05$) spatial patterns: High Crash-High HighG (HH), Low Crash-Low HighG (LL), High Crash-Low HighG (HL), and Low Crash-High HighG (LH), revealing areas of concordance and discordance between the two indicators. 
\end{itemize}

Following the identification of distinct spatial correlation patterns via LISA, a characterization focused on the environmental context provided by Points of Interest.

POIs joined with the spatial grid in this study originate from two primary sources: OpenStreetMap (OSM) amenity data and a dedicated traffic light dataset provided by Transport for NSW.

\begin{figure*}[htbp] 
\centering
\includegraphics[width=0.5\linewidth]{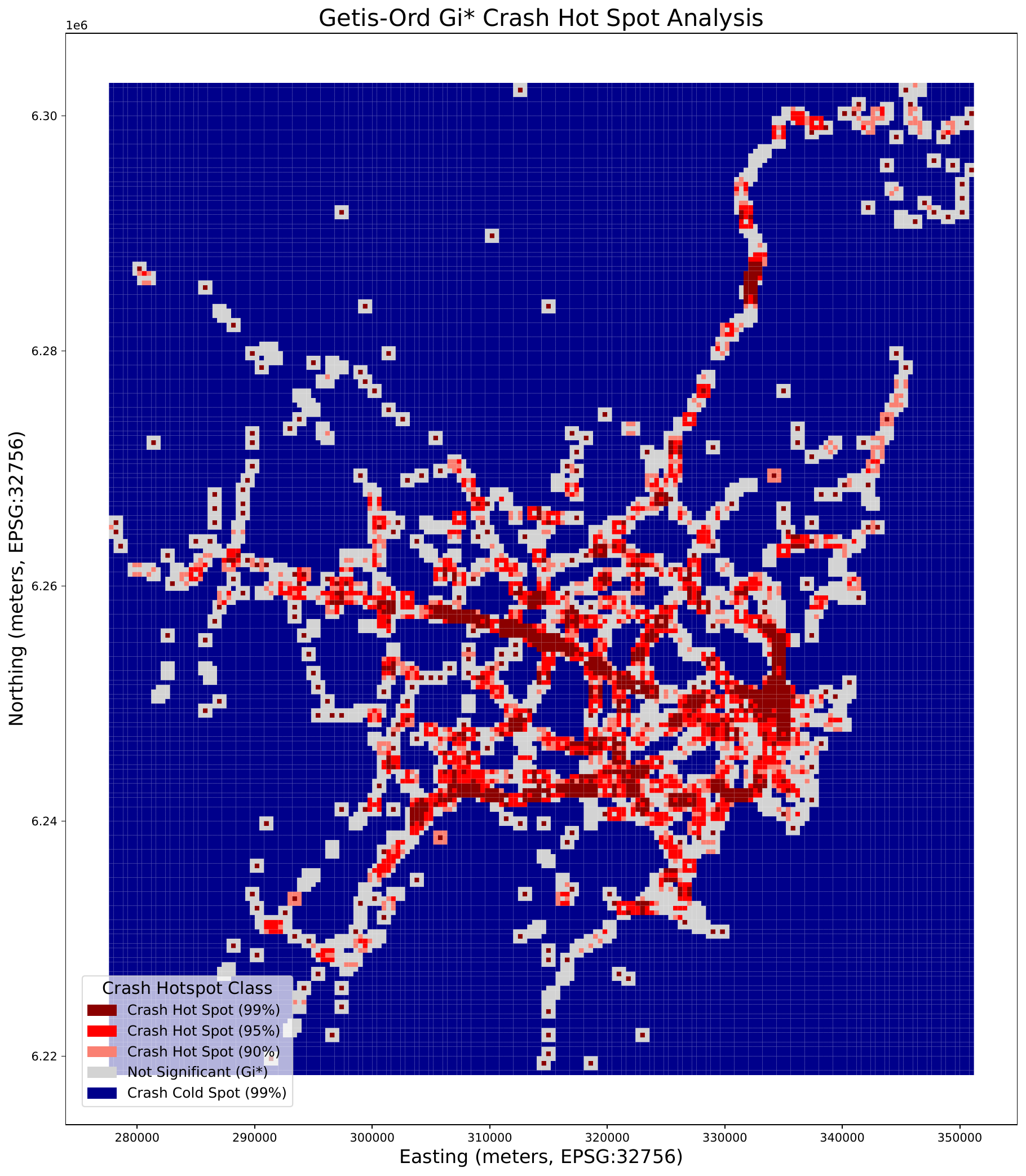} 
\caption{Spatial distribution of statistically significant \textbf{road accidents} clusters on the \textbf{400m grid}, based on Getis-Ord $G_i^*$ analysis for 2022. Red/salmon areas indicate hotspots (significantly high concentration), blue areas indicate coldspots (significantly low concentration), and grey areas represent locations with no statistically significant clustering at the $p < 0.05$ level.} 
\label{fig:crash_hotspot_map} 
\end{figure*}

The Getis-Ord $G_i^*$ statistic was used to identify statistically significant spatial clusters of reported crashes (A) across the 400m grid cells covering the Sydney study area (circa 2022). This identifies localized concentrations significantly higher (hotspots) or lower (coldspots) than expected given the overall spatial distribution. The resulting spatial pattern of crash hotspots and coldspots is illustrated in Fig. \ref{fig:crash_hotspot_map}.

Getis-Ord $G_i^*$ excels at identifying statistically significant local clusters of hot or cold spots precisely because it explicitly accounts for spatial autocorrelation by considering the values of neighboring features. In contrast, Kernel Density Estimation (KDE) primarily focuses on estimating the spatial density of point locations and typically assumes an isotropic (uniform in all directions) spread, without accounting for spatial autocorrelation.

\begin{figure*}[htbp]
\centering

\includegraphics[width=0.5\linewidth]{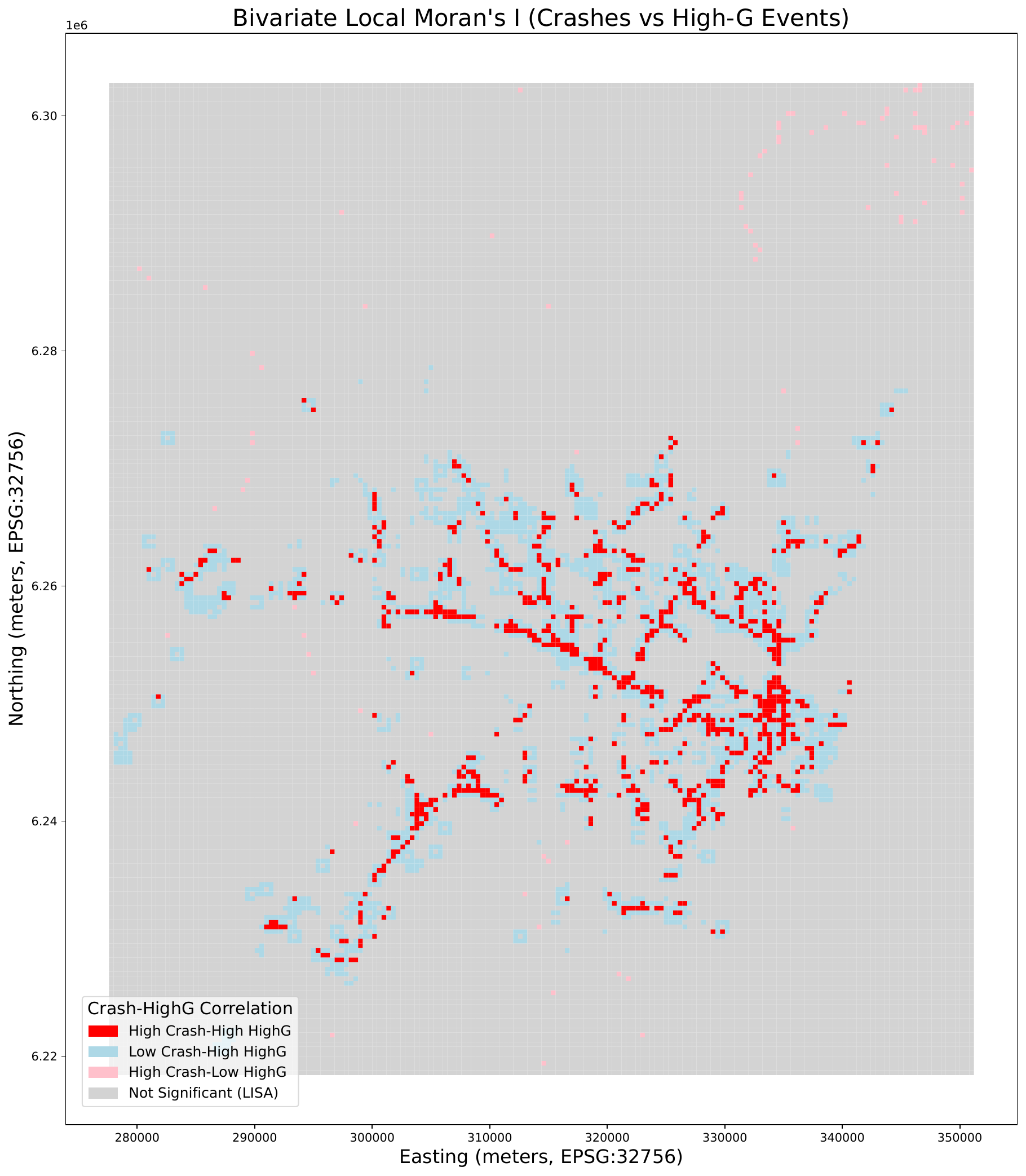} 
\caption{Map of LISA results on the \textbf{400m grid}, showing the spatial correlation between \textbf{crash counts} and \textbf{High-G event counts} per cell. Colors represent the type of statistically significant ($p < 0.05$) local correlation: High-High (Red: high crashes, high High-G), Low-Low (Blue: low crashes, low High-G), High-Low (Pink: high crashes, low High-G), Low-High (Light Blue: low crashes, high High-G). Grey areas indicate no significant local spatial correlation. (Note: Analysis log saved this plot as crash\_highg\_correlation.pdf)} 
\label{fig:crash_highg_correlation_map} 
\end{figure*}

\textbf{Local Spatial Correlation between Crash Counts and High-G Event Counts:} To investigate the local relationship between the frequency of reported crashes and the frequency of high G-force (High-G) events, a LISA analysis was performed on the 400m grid cell data. This analysis assesses whether the spatial pattern of \textbf{crash counts} in a given cell is significantly correlated with the spatial pattern of \textbf{High-G event counts} in its neighborhood (using Queen contiguity). The geographical distribution of these local spatial correlations is depicted in \ref{fig:crash_highg_correlation_map}. 

\section{Results}


The spatial distribution and correlation of reported crashes ($n=\num{3638}$) and high G-force (High-G) events ($n=\num{24137}$) were analyzed using a 400m\,x\,400m square grid resolution, encompassing \num{38824} cells across the Sydney study area. Initial analysis using Global Moran's I confirmed significant positive spatial autocorrelation (clustering) for both crash counts (Moran's I = \num{0.2080}, $p=0.001$) and High-G event counts (Moran's I = \num{0.2586}, $p=0.001$). This indicates an overall tendency for both phenomena to cluster spatially rather than being randomly distributed.

Furthermore, the overall spatial relationship between the two variables was assessed using Global Bivariate Moran's I, which revealed a significant positive association (Global Bivariate Moran's I = \num{0.1654}, $p=0.001$). This suggests that, on average, areas with high crash counts tend to be spatially close to areas with high High-G event counts across the study region.

\subsection{Identification and Classification of Near-Miss Hotspots}

This analysis investigates the spatial relationship between potential traffic conflicts, represented by high G-force events (High-G, $n=24,137$), and realized harm, represented by reported crash locations (Crashes, $n=3,638$). The study area was divided into a regular grid with 400m x 400m cells.

To understand the spatial concordance or discordance between these indicators, a LISA analysis was conducted. This spatially correlates the counts of High-G events with the counts of reported Crashes within the same grid cell system (using Queen contiguity weights). The analysis identifies locations where the local concentration of High-G events significantly aligns (or fails to align) with the local concentration of Crashes. Based on the LISA results (significant at $p < 0.05$), grid cells were classified into distinct spatial relationship categories. These classifications reveal distinct spatial patterns comparing crash counts and high G-force (High-G) event counts within 400m grid cells (Table~\ref{tab:lisa_groups_results}):

\begin{itemize}
    \item \textbf{HH (High Crash-High High-G):} Represents 825 grid cells where high counts of potential conflicts (High-G events) spatially coincide with high counts of realized harm (Crashes). These areas strongly indicate locations where risky driving maneuvers are frequent and may directly translate into reported crashes. These are the primary co-located hotspots.
    \item \textbf{HL (High Crash-Low High-G):} Represents 91 grid cells with historically high crash counts that are not statistically associated with high counts of the measured High-G events in their vicinity. This suggests that crash risk in these locations might be driven by factors other than those frequently captured by the High-G metric (e.g., specific static hazards, safety-focused intersection designs).
    \item \textbf{LH (Low Crash-High High-G):} Represents a substantial number of areas (2681 grid cells) characterized by statistically significant high counts of potential conflicts (High-G events) but low counts of actual reported crashes. These locations with frequent near-misses but low crash history warrant proactive monitoring. They suggest the presence of frequent risky maneuvers or situations, but perhaps mitigating factors (e.g., effective road design allowing recovery, lower speeds, successful evasive actions, under-reporting of minor crashes) currently prevent these from translating into high numbers of reported crashes. These areas warrant proactive monitoring.
    \item \textbf{LL (Low Crash-Low High-G):} Represents areas where low potential conflict (High-G) and low realized harm (Crashes) would theoretically coincide, likely indicating baseline safer conditions or lower exposure areas. However, the analysis found no grid cells (0 cells) exhibiting a statistically significant Low-Low spatial relationship at the $p < 0.05$ level. While many areas likely have low counts of both, they don't form statistically significant spatial clusters of low-low association according to this specific bivariate test.
    \item \textbf{Not Significant:} The majority of grid cells (35227) did not show a statistically significant spatial correlation ($p \ge 0.05$) between local crash counts and neighbouring High-G counts (or vice-versa) under the Bivariate LISA test.
\end{itemize}

\begin{table*}[htbp]
\scriptsize
\centering
\caption{Classification and Size of Grid Cells Based on Bivariate Spatial Correlation (LISA, p < 0.05) between Crash Counts and High G-Force Event Counts.}
\label{tab:lisa_groups_results} 
\begin{tabular}{@{}llc@{}}
\toprule
Spatial Relationship         & LISA Classification          & Number of Cells (n) \\
\midrule
High Crash \& High High-G    & HH (High Crash-High HighG) & 825 \\
High Crash \& Low High-G     & HL (High Crash-Low HighG)  & 91  \\
Low Crash \& High High-G     & LH (Low Crash-High HighG)  & 2681 \\
Low Crash \& Low High-G      & LL (Low Crash-Low HighG)   & 0\textsuperscript{*}   \\ 
Not Spatially Correlated     & Not Significant (LISA)     & 35227 \\ 
\bottomrule
\multicolumn{3}{@{}l@{}}{\textsuperscript{*} \footnotesize{No grid cells showed a statistically significant Low-Low spatial relationship at $p < 0.05$.}} \\ 
\end{tabular}
\end{table*}

These local patterns highlight specific areas of concordance (HH) and discordance (HL, LH) between historical crash data and the High-G near-miss indicator.
Comparing these different spatial clusters, particularly the high number of LH cells (2681) versus HH cells (825), suggests that while High-G events are spatially associated with crashes in many hotspots (HH), there are even more areas where frequent near-misses occur without a corresponding high crash history (LH). This highlights the potential of High-G data for proactive safety analysis, identifying areas of concern before they become crash blackspots, and potentially revealing locations where safety interventions or specific road characteristics are effectively mitigating crash outcomes despite frequent risky events. Further analysis of the characteristics differentiating HH, HL, and LH areas is crucial.


Findings from the Mann-Whitney U tests conducted to compare the prevalence of various Points of Interest (POIs) between grid cells classified as 'High Crash-High HighG' (Clusters) and those classified as 'Low Crash-Low HighG' (Outliers) (see Table \ref{tab:mw_results}). The tests aimed to identify statistically significant differences in the distribution of POI counts between these two area types, using a significance level ($\alpha$) of 0.05.

\begin{table*}[htbp] 
\scriptsize
  \centering
  \caption{Mann-Whitney U Test Results: POI Counts in High Crash-High HighG vs Low Crash-Low HighG Areas}
  \label{tab:mw_results}
  \begin{tabular}{l r r r r c}
    \toprule
    POI Type & U Statistic & p-value & Mean Outliers (LL) & Mean Clusters (HH) & Significant ($\alpha=0.05$) \\
    \midrule
    appliance      & 1109934.0 & 0.001791 & 0.000000 & 0.003636 & True \\
    bed            & 1109934.0 & 0.001791 & 0.000000 & 0.006061 & True \\
    viewpoint      & 1122576.0 & 0.001928 & 0.014174 & 0.027879 & True \\
    radiotechnics  & 1108593.5 & 0.010795 & 0.000000 & 0.002424 & True \\
    flooring       & 1108593.5 & 0.010795 & 0.000000 & 0.002424 & True \\
    trophy         & 1108593.5 & 0.010795 & 0.000000 & 0.002424 & True \\
    lighting       & 1108593.5 & 0.010795 & 0.000000 & 0.002424 & True \\
    prep\_school   & 1108593.5 & 0.010795 & 0.000000 & 0.002424 & True \\ 
    baby\_goods    & 1108593.5 & 0.010795 & 0.000000 & 0.002424 & True \\ 
    fabric         & 1110449.5 & 0.012690 & 0.000746 & 0.004848 & True \\
    health\_food   & 1109521.5 & 0.015208 & 0.000373 & 0.003636 & True \\ 
    pawnbroker     & 1109521.5 & 0.015208 & 0.000373 & 0.003636 & True \\
    traffic\_signals & 1076733.0 & 0.017472 & 0.441999 & 0.303030 & True \\ 
    waste\_basket  & 1123232.0 & 0.022249 & 0.082432 & 0.136970 & True \\ 
    bench          & 1128707.0 & 0.030349 & 0.303991 & 0.449697 & True \\
    \bottomrule
  \end{tabular}
\end{table*}

\textbf{Key Observations:}

\begin{enumerate}
    \item \textbf{Distinct POI Profiles:} The results strongly indicate that the environmental characteristics, as represented by POI types, differ significantly between the High Crash-High HighG and Low Crash-Low HighG areas.

    \item \textbf{POIs More Prevalent in High Crash-High HighG Areas:} A notable number of POI types were found to be significantly more common in the High-High cluster areas: retail and commercial services (e.g., appliances, lighting), public amenities (benches, viewpoints), and educational facilities. The higher density of benches and viewpoints might suggest areas with higher pedestrian activity.

    \item \textbf{POIs Less Prevalent in High Crash-High HighG Areas:} Perhaps the most striking finding in this category is the significantly lower prevalence of traffic signals in High-High areas compared to Low-Low areas (p=0.017). This suggests that the High-High areas, despite having more crashes and High-G events, might be less characterized by major, signalized intersections compared to the Low-Low areas used in this comparison. This could point towards differences in road network hierarchy or potentially higher prevalence of unsignalized intersections or different road types (e.g., mid-block segments) contributing to the High-High classification.

    \item Several types of points of interest, including music schools, safety equipment suppliers, arts centers and tea shops, did not show a statistically significant difference between the two groups. These points of interest were generally rare in both area types, as indicated by their low average counts.
\end{enumerate}

Infrastructure improvements (e.g., signal timing optimization, pedestrian crosswalks) should be prioritized in HH areas given their dual risk of crashes and near-misses. For LH areas, extending monitoring systems for proactive conflict detection, as these sites may evolve into crash blackspots.

\section{Conclusion}
\label{sec:conclusion}

This study investigated the spatial relationship between reported road crashes and a high-severity near-miss proxy (High-G events) in Sydney using LISA over a 400m grid. The analysis confirmed significant spatial clustering of crashes and revealed distinct local patterns of association between the two indicators. Concordant hotspot areas were identified where high crash rates coincided with statistically high neighbouring High-G rates (HH pattern, \num{825} cells). However, no areas showed a statistically significant pattern where both indicators were concurrently low (LL pattern, \num{0} significant cells found at $p < 0.05$). Crucially, discordant areas were also found: locations with high crashes but statistically low neighbouring High-G events (HL pattern, \num{91} cells), and a substantial number of locations (\num{2681} cells) with low crashes despite statistically high neighbouring High-G events (LH pattern).

These findings demonstrate that the spatial distribution of High-G events shows statistically significant concordance (p<0.05) with historical crash patterns in \num{825} cells (HH clusters), while identifying \num{2681} additional high-risk areas (LH clusters) where frequent near-misses occur without corresponding crashes. This indicates High-G data identifies additional locations requiring monitoring beyond those identified solely by historical crash data, potentially revealing areas where frequent risky maneuvers occur without corresponding crash history.


Characterizing the different pattern areas using POI data revealed significant differences in land use context between these spatial patterns. The discordant patterns are particularly valuable: HL areas (\num{91}) may point to crash causes not captured well by the High-G metric, while the numerous LH areas (\num{2681}) suggest locations with frequent risky events where crashes are currently mitigated or under-reported, warranting proactive investigation and monitoring. Limitations include data accuracy, the specific definition and proxy nature of High-G events. Future work should incorporate dynamic traffic flow and detailed infrastructure data to better explain these complex spatial patterns and advance proactive road safety strategies.

Future research should focus on incorporating dynamic traffic variables, detailed road infrastructure data, and potentially localized driver behaviour information to build more comprehensive models explaining the observed spatial patterns of concordance and discordance. Network-based analysis and qualitative case studies of specific HL and LH locations could provide further findings into the factors mitigating or exacerbating risk. Ultimately, an approach combining leading and lagging indicators with contextual data is essential for advancing proactive road safety management.

\section{ACKNOWLEDGEMENTS}
We thank Compass IoT for the data and support provided for this study. This work has been funded by the UTS Jenny Edwards Fellowship awarded in 2025 to Assoc. Prof. Adriana-Simona Mihaita for conducting research in risky driver behaviour identification.

\bibliographystyle{IEEEtran} 
\bibliography{references, BIBLIO_Simona_2025_April}       

\end{document}